\documentclass{article}
\usepackage[a4paper,
            bindingoffset=0.2in,
            left=1.2in,
            right=1.2in,
            top=1in,
            bottom=1in,
            footskip=.25in]{geometry}

\usepackage{lipsum}
\usepackage{amsfonts}
\usepackage{amsopn}
\usepackage{graphicx}
\usepackage{epstopdf}
\usepackage{algorithmic}
\usepackage{bm}
\usepackage{amsmath}
\usepackage{cleveref}
\pdfoutput=1 
\usepackage{xcolor}
\usepackage{lscape}
\usepackage{subfig}
\usepackage[normalem]{ulem}
\usepackage{array}

\usepackage{enumitem}
\setlist[enumerate]{leftmargin=.5in}
\setlist[itemize]{leftmargin=.5in}

\title{Modeling Stock Return Distributions and Pricing Options}

\author{Xinxin Jiang\thanks{Mathematics and Computer Science Department, Suffolk University. (xjiang@suffolk.edu).}}

\begin{document}
\maketitle

\begin{abstract}
{ This paper provides evidence that stock returns, after truncation, might be modeled by a special type of continuous mixtures or normals, so-called $q$-Gaussians. Negative binomial distributions might model the counts for extreme returns. A generalized jump-diffusion model is proposed, and an explicit option pricing formula is obtained.}
\medskip

\end{abstract}

\section{Introduction}

The celebrated Black and Scholes option pricing formula \cite{Black1973} is obtained based on several basic assumptions. One of those is that the logarithmic returns of stock prices follow a Brownian Motion. It is also well known that the empirical log-returns show some stylized facts, such as thick tails, volatility clustering, leverage effects, information arrivals, long memory and persistence, volatility comovements, implied volatility correlations, the term structure of implied volatilities, and smiles \cite{Ghysels1996}. To address some of these issues, numerous generalizations have been proposed. One of them is the Merton jump diffusion model \cite{Merton1976}. The model assumes that the stock price $S_t$ follows the random process \footnote{The SDE presented here adopts the notation from an online source \cite{Peter2011}. It appears differently from Merton's original SDE, but they are essentially the same.}
$$ \frac{dS_t}{S_t}=\mu dt+ \sigma dB_t+(J-1)dN(\lambda t). \eqno(1)$$ The first two parts are from the classic Black-Scholes model, where $\mu$ is the drift rate, $\sigma$ is the constant volatility, and $B_t$ is the Brownian motion. The last term represents the jumps, where $N(\lambda t)$ is a Poisson process representing the number of jumps up to time $t$ with parameter $\lambda t$. $\lambda$ is a constant here. $J$ is the jump size as a multiple of the stock price with distribution commonly assumed to be log-normal. \footnote{\cite{CaiKou2011} proposes a mixed-exponential distribution to model the jump size, which was not considered in this paper.} The Brownian motion and the Poisson process are assumed to be independent. 
\par
For the continuous part, \cite{Borland2002} uses $q$-Gaussians and a nonlinear Fokker-Plank equation to model log-return distributions and movements. \cite{Hahn2010} views $q$-Gaussians as variance mixtures of normals and proposes an exchangeable process to model log-return movements.
\par
We generalized Merton's jump-diffusion model in two ways. First, the Brownian motion is replaced by more general variance mixtures of Brownian motions. That is, the $\sigma$ in (1) will be treated as a random variable rather than a constant. Second, the Poisson process is replaced by a negative binomial process, which may be viewed as a mixture of Poisson processes with $\lambda$ selected from a Gamma distribution.
\par
The remaining part of the paper is organized as follows. In Section 2, using S\&P 500 historical data, evidence is shown that variance mixtures of normals might be appropriate to model logarithmic returns after truncation. For the number of extreme log returns (viewed as jumps), a negative binomial distribution might be preferable to a Poisson distribution. In Section 3, a generalized jump-diffusion model is proposed and an explicit formula for option prices is given. Section 4 presents empirical results. Section 5 discusses and concludes.

\section {Empirical analysis on the distribution of stock daily returns}

\subsection {Preliminary results}
\par
 In this subsection, we will use S\&P 500 as an example to analyze the distribution of daily returns\footnote{We did similar work for Dow and Nasdaq. The results were similar}. Data were downloaded from two resources: WRDS (Wharton Research Data Services, through Suffolk University) and finance.yahoo.com. There are 24029 records of closing points from 1928/1/3 to 2023/9/1. Denote them by $S_n$. Then the daily returns are calculated by $DR_n=\ln S_{n+1}-\ln S_n$.
\par
 The history of S\&P 500 dates back to 1923 when 233 companies were covered. The index as it is known today started in 1957 and has been covering about 500 companies since then. We are unable to get the data back to 1923. We do include data before 1957 to have more records to analyze.
\par
 Since there are on average about 252 trading days, $23940=252\times 95$ returns are selected and the most recent data are discarded. The number of years used is 95.
\par
  There are 313 zeros, representing 1.3\% of the data. It might be debatable whether we should treat them differently. On the one hand, if we model the daily returns by a continuous distribution, the probability of a daily return taking a value of zero is zero. However, the closing points are discrete anyway, which makes it possible to have multiple values in practice. We will include zeros in our analysis without treating them separately.
  \par
  For each year of 252 records, we first break them into two groups. Those in $[Q1-1.5*IQR, Q3+1.5*IQR]$ are called truncated returns, and the others are outliers. Here, $Q1$ and $Q3$ are the first quartile and the third quartile, respectively. $IQR$ is the interquartile range. This criterion is often used in elementary statistics.
\par
 The intuition behind it is that daily returns have some extremely large values. It is then difficult to use one continuous distribution to model all the return data. It would be better to treat them separately. In this paper, these returns are treated as jumps.
\par
\begin{figure}
\centering
    \includegraphics[scale=.4]{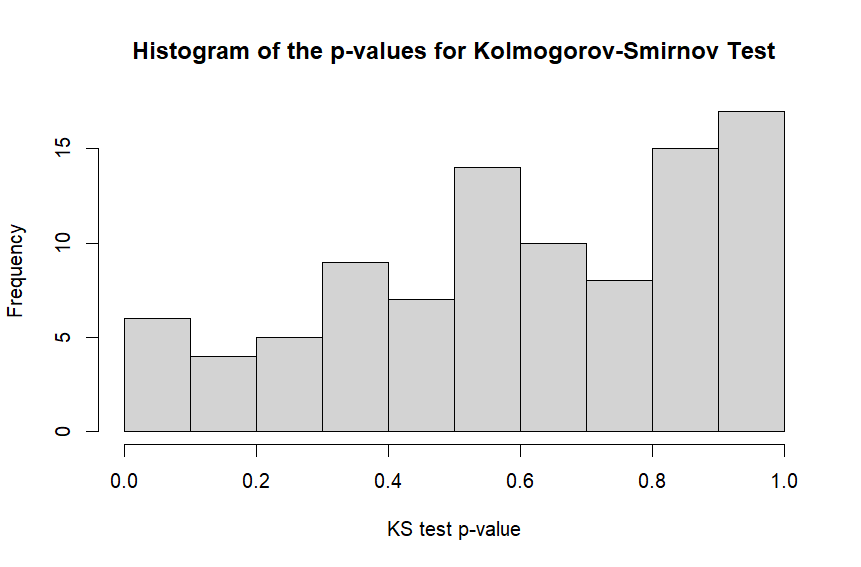}
    \includegraphics[scale=0.4]{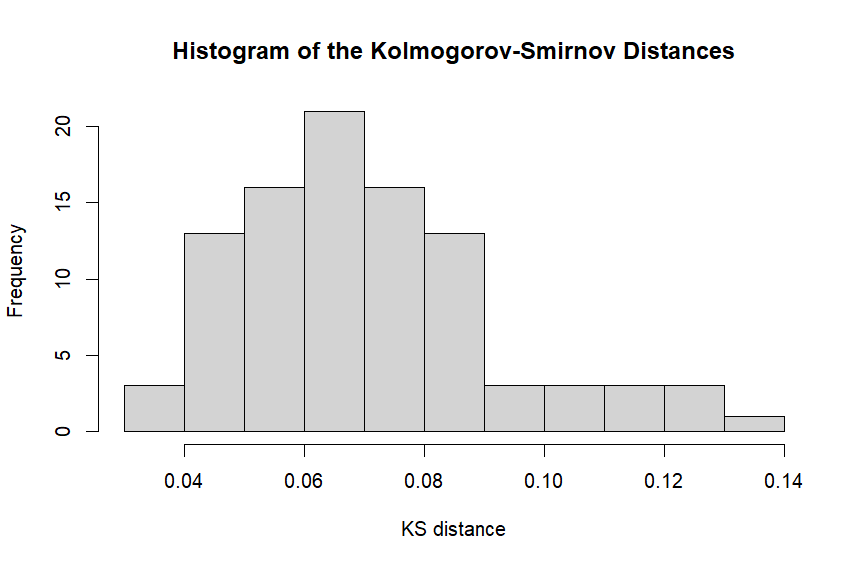}
    \caption{These results show that the Kolmogorov-Smirnov test cannot reject the null hypothesis that the daily returns after being truncated follow a normal distribution for most of the years.}
    \label{hist_kstest}
\end{figure}
\begin{figure}
\centering
    \includegraphics[scale=.25]{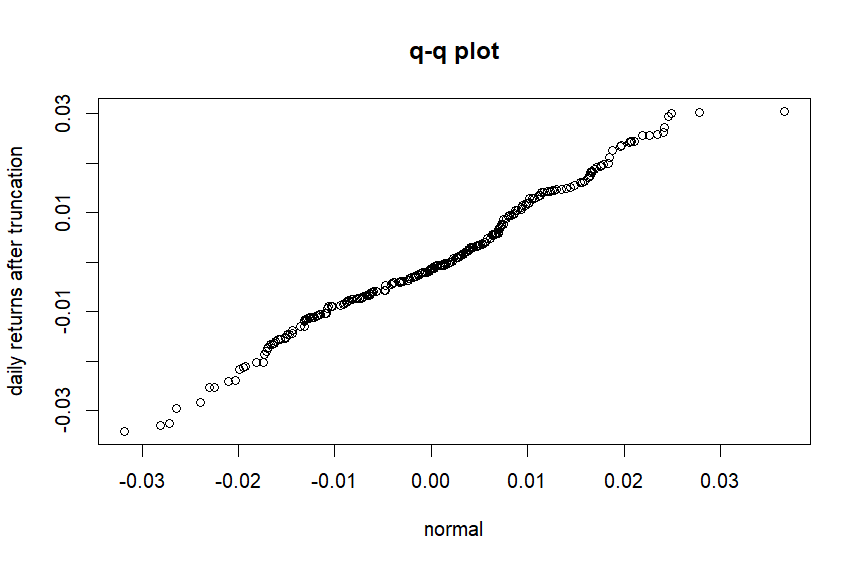}
    \includegraphics[scale=.25]{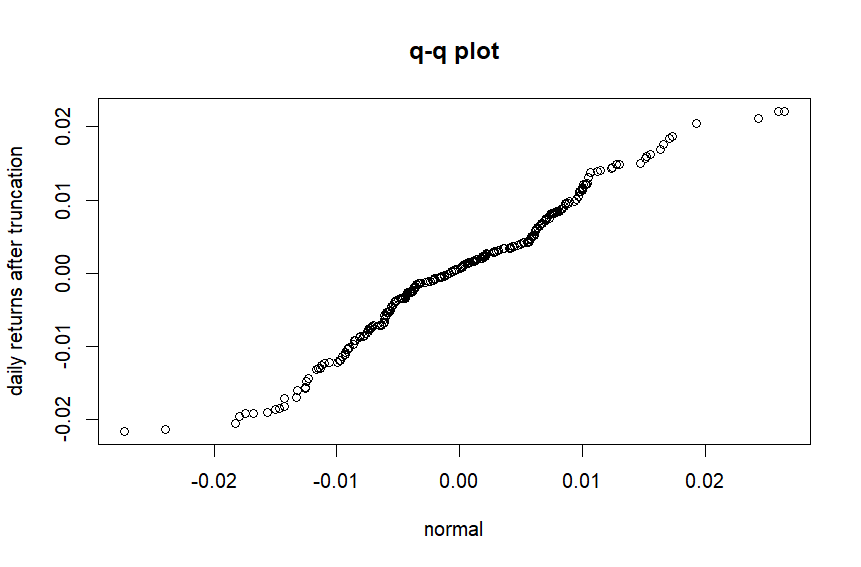}
    \includegraphics[scale=.25]{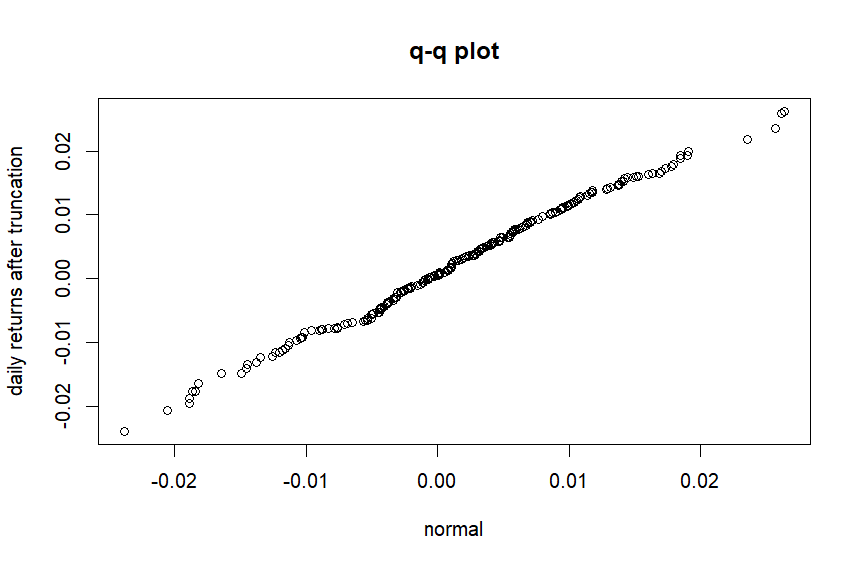}
    \includegraphics[scale=.25]{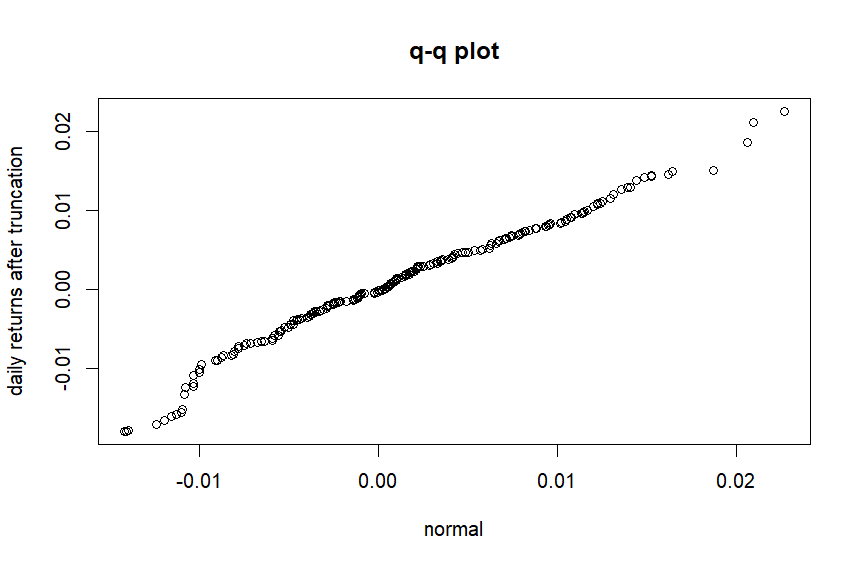}
    \includegraphics[scale=.25]{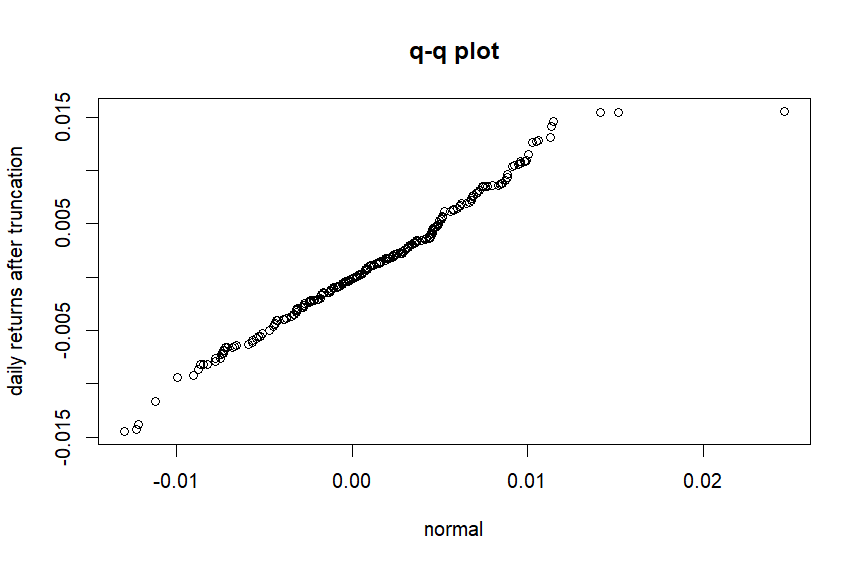}
    \includegraphics[scale=.25]{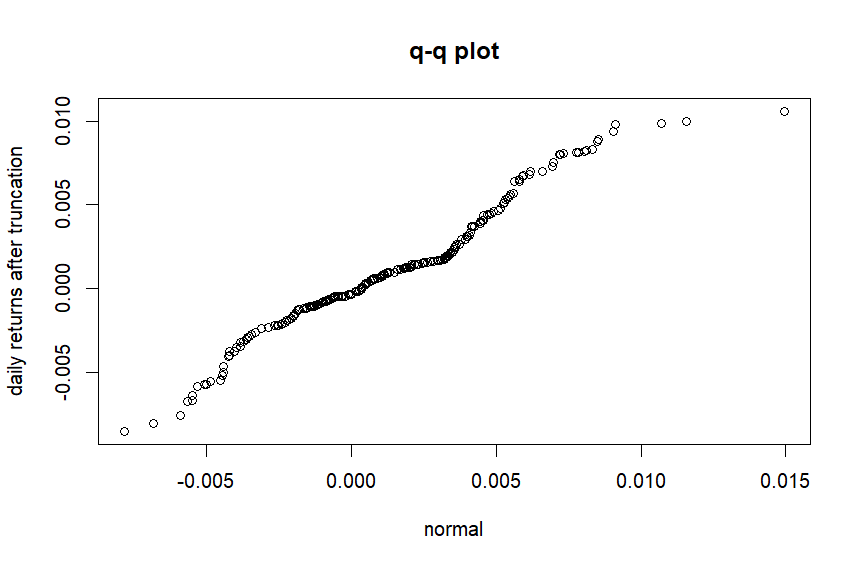}
    \caption{q-q plots for the last six years are presented here. Mostly they show some linearity.}
    \label{qqplots}
\end{figure}
\begin{figure}
\centering
 \includegraphics[scale=.4]{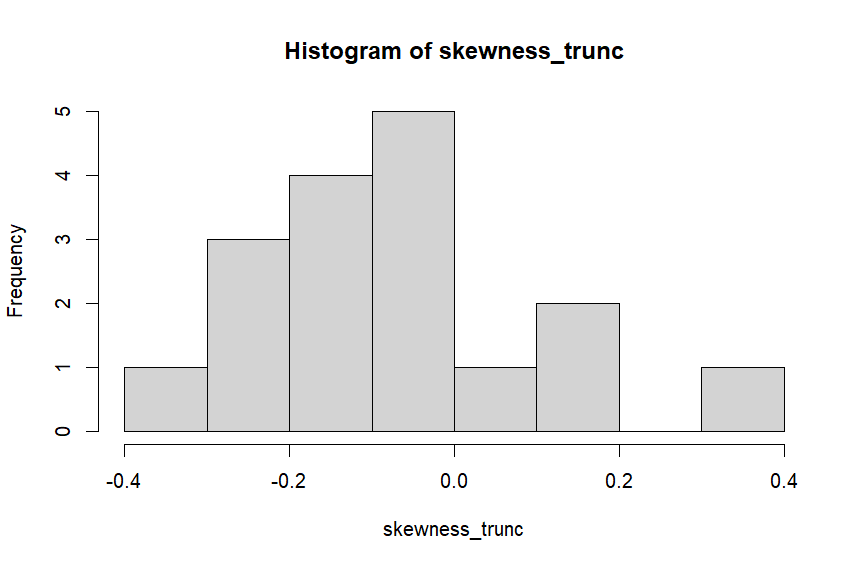}
  \includegraphics[scale=.4]{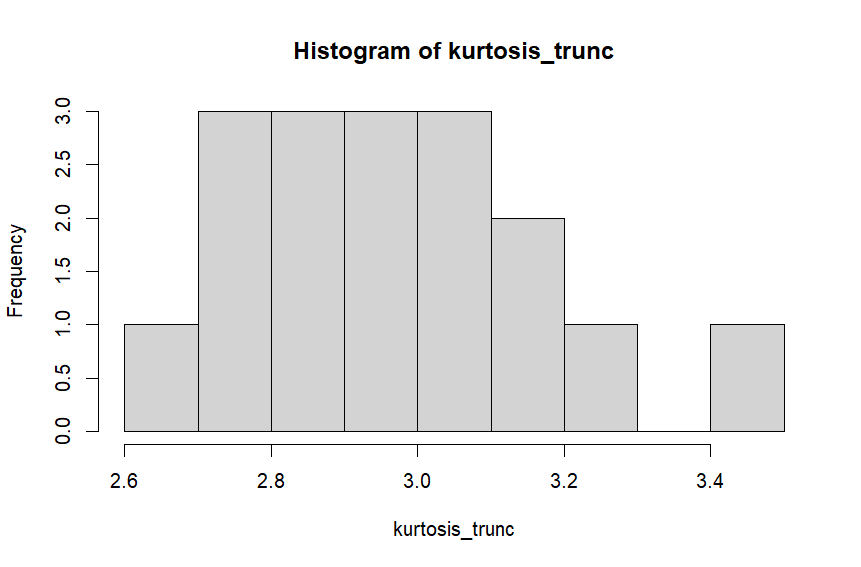}
  \caption{The left is the histogram for the skewness of the truncated daily returns with mean= -0.0375, which confirms that the returns are skewed to the left generally. The right is for the kurtosis with mean=2.88. More kurtoses are less than 3, possibly due to the truncation effect.}
  \label{skewness_kurtosis}
\end{figure}

For truncated returns, $q-q$ normality plots are drawn and Kolmogorov-Smirnov normality tests are run.  Histograms of the $p$-values and Kolmogorov-Smirnov distances are shown in \cref{hist_kstest}. Some of the $q-q$ plots are reported in \cref{qqplots}. \cref{skewness_kurtosis} shows the histograms of skewness and kurtosis.

\par

 The results show that for most years, the $q-q$ plots are close to lines which shows some evidence of normality for truncated returns. Only about 5\% of the $p$ values are less than 0.1. Although these results will not prove that truncated returns follow normal distributions, large $p$ values, and small Kolmogorov-Smirnov distances do give some evidence that normal distributions seem to be good candidates to model truncated returns annually. \footnote{Shapiro-Wilk test is arguably a better choice (more powerful test) to test normality. We run the test and, indeed, the $p$-values are generally smaller, 14 out of 95 less than 5\% and 23 less than 10\%.}

\begin{figure}
\centering
    \includegraphics[scale=.5]{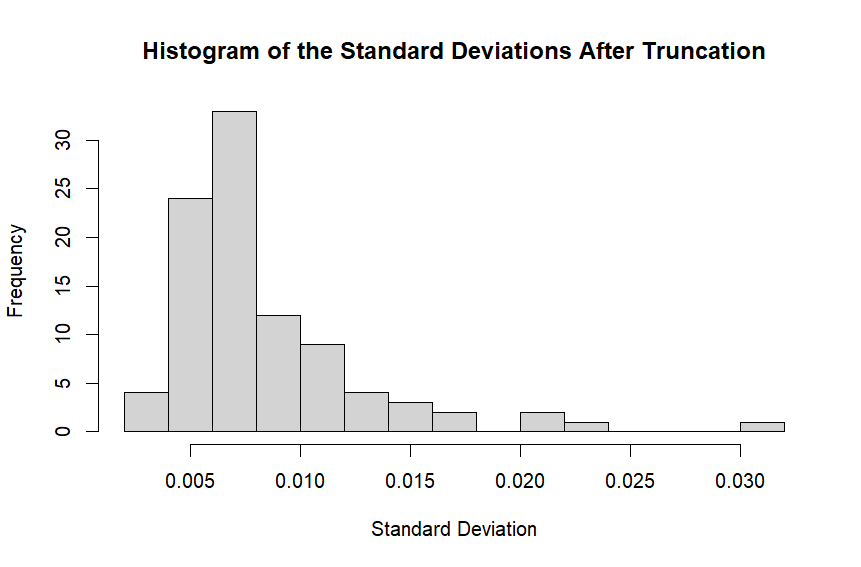}
        \caption{The standard deviations are daily for each year. There are 95 records.}
    \label{hist_sd}
\end{figure}

Next, we find the standard deviation for the truncated returns each year. The histogram is shown in \cref{hist_sd}. As expected, the standard deviations are quite spread. $F$-tests are run for the ratios of these 95 standard deviations. Within 4465 ratios, the largest 1000 ratios give $p$-values of the $F$-test less than 0.01. This implies that these pairs of variances are not considered the same. However, when the years are close, the variances are generally close.

These results suggest that the truncated daily returns for S\&P 500 may be modeled by a variance mixture of normals (VMON).

Remarks:

1. Retrospectively, it's worth pointing out that we hypothesize that the truncated daily returns follow a VMON distribution. There is no guarantee that the variance within each year is constant. It is then particularly striking that the strong linearity is shown by the $q-q$ plots. This might imply that it's probably safe to assume constant standard deviation in a short period such as a year.

\begin{figure}
\centering
    \includegraphics[scale=.5]{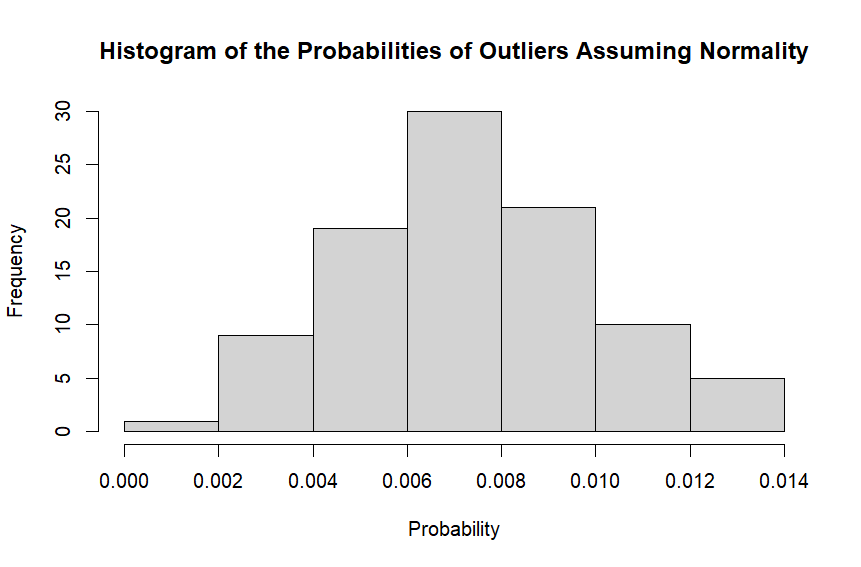}
        \caption{x-axis represents the probability that a value is beyond the benchmark. That is the probability a data point is considered to be an outlier. There are 1001 records. The majority of probabilities are less than 0.01, which may suggest that the benchmark works well. }
    \label{prob_outliers}
\end{figure}

2. To check whether the benchmark $[Q1-1.5*IQR, Q3+1.5*IQR]$ is reasonable, we calculate the probabilities outside the intervals using a normal distribution with mean and standard deviation from the truncated returns, annually. The histogram of these probabilities is shown in \cref{prob_outliers}. The majority of the probabilities are less than 1\%.

\par

\begin{figure}
\centering
    \includegraphics[scale=.4]{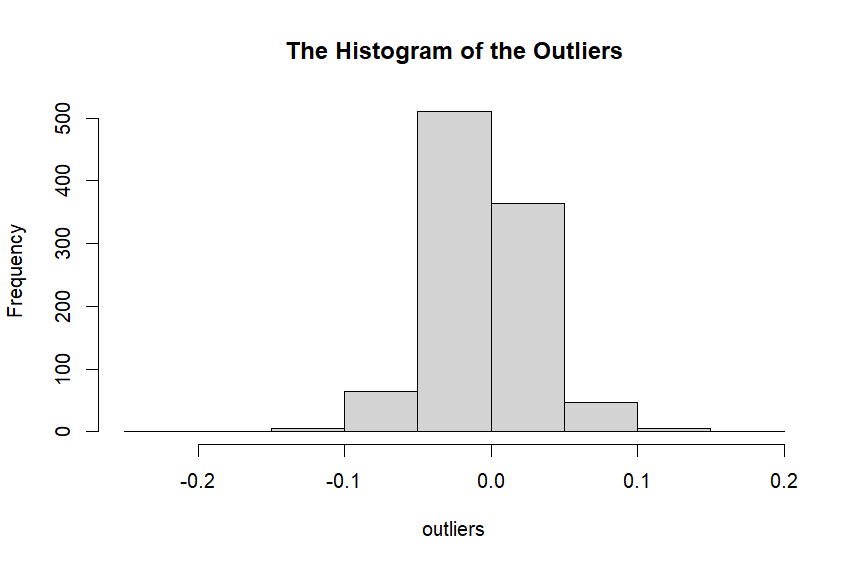}
    \includegraphics[scale=.4]{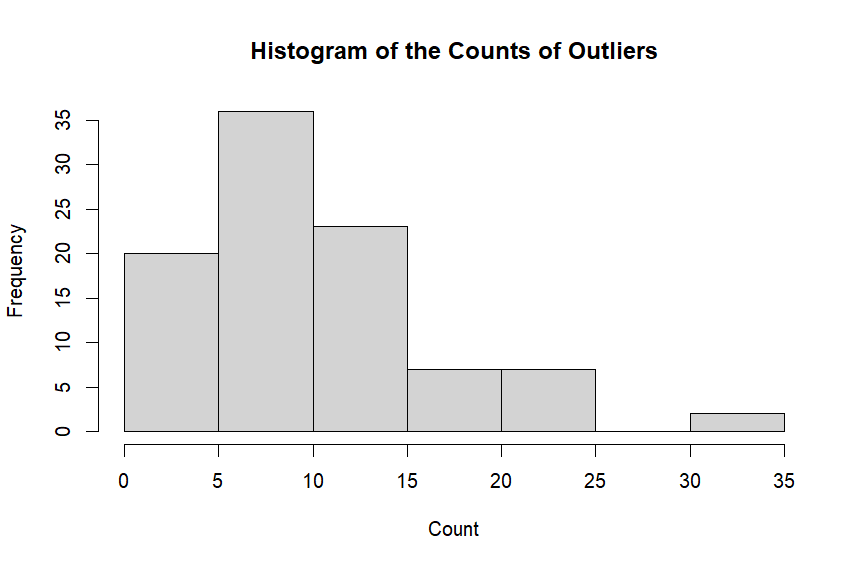}
        \caption{The left is the distribution of the outliers. $x$ values are extreme returns. The right side is the distribution of the counts of the outliers.}
    \label{outlierdist}
\end{figure}

 Let us turn our attention to the outliers. There are 1001 records, about 5\% of the total returns. The histogram is shown on the left side of \cref{outlierdist}. Their annual counts with a total of 95 records are shown on the right side of \cref{outlierdist}.
 
\par
 The distribution to model the counts is usually assumed to be Poisson because extreme returns are considered to be rare events. This is debatable. For example, when we calculate the mean (10.5) and variance (43.2) of the counts, they are not equal. The behavior is consistent even if we widen the range of the benchmark to include fewer extreme returns (so the returns included in outliers will presumably be closer to Poisson events).
\par
An alternative distribution that captures the fact, that is, variance$>$mean, sometimes called overdispersion, is negative binomial. We will discuss it further shortly.
\par
It is probably worthwhile looking into how extreme returns would happen in reality. Some of them might be caused by unanticipated events such as 911 or COVID-19, which might be better modeled by Poisson. The overdispersion might be caused by clustering, meaning that their effect would not be finished in one or two days. Some other of them might be caused by anticipated events such as Feds rate decision, earning releasing, important data releasing, etc. The Poisson model might not be appropriate even theoretically.
\par
In this paper, however, we will treat all 1001 records as extreme events and analyze them all together.

\subsection {Truncated returns and variance mixture of normals (VMON)}

\par
Variance mixtures of normals include a wide range of distributions such as stable distributions and student $t$ distributions. They, together with discrete VMON distributions, were used to model stock returns. In this subsection, the so-called Tsallis $q$-Gaussians will be used to model the truncated returns. They have been widely discussed in recent years because they have shown some universality as limiting distributions in areas such as anomalous diffusion in type-II superconductors, granular matter, non-Gaussian momenta distributions for cold atoms in optical lattices, dirty plasma, trapped atoms, etc. \cite{Umarov_2016}. Their applications in finance were studied in \cite{Borland2002}, \cite{Iliopoulos2015}, among others. $q$-Gaussians, when $1<q<3$, which is the case we are considering here, are variance mixtures of normlas (VMON)\cite{Hahn2010}. The density is given by

$$G_q(\beta;z)={\sqrt\beta\over C_q}e_q^{-\beta z^2},$$ with the norming constant $$C_q={\sqrt\pi\Gamma({3-q\over 2(q-1)})\over \sqrt{q-1}\Gamma({1\over q-1})},$$ and $e_q(x)=[1-(1-q)x]_+^{1/(1-q)}$. Here, $\beta$ is the scaling parameter. The relation between $\beta$ and $q$ is given by $$variance*\beta *(5-3q)=1$$ for $q<5/3$, which is sufficient for our purposes.

 When $q=1$, $q$-Gaussian becomes normal. In \cite{Beck2001}, the $q$-Gaussians are identified as ${1\over a}Z$ where $Z$ is a standard normal random variable and $a^2$ has a $\chi^2$ distribution with $q$ derived from the degrees of freedom. $Z$ and $a$ are independent of each other. It is also known that $q$-Gaussians are scaled reparameterization of $t$ distributions.

To use $q$-Gaussians, it is important to estimate $q$. It seems that there is no nice method (such as the maximum likelihood method) estimating $q$. \cite{Ferri2010} uses the fact that $\ln_q(e_q^{-\beta z^2})$ is a linear function with respect to $z^2$, and obtains a method to estimating $q$ by maximizing the correlation coefficient between $\ln_q(e_q^{-\beta z^2})$ and $z^2$. Here $\ln_q(x)={x^{1-q}-1\over 1-q}$ with $\ln_1(x)=\ln(x)$. So $$\ln_q[e_q(x)]=e_q[\ln_q(x)]=1.$$

In this paper, we use the fact that the variance of a $q$-Gaussian is $\displaystyle {1\over \beta (5-3q)}$ when $q<5/3$ and estimate $q$ by maximizing the sum of the differences between a $q$-Gaussian and the empirical distribution (after centering). 313 zero returns (representing 1.36\% of 22939 truncated returns) are discarded.

\begin{figure}
\centering
    \includegraphics[scale=.5]{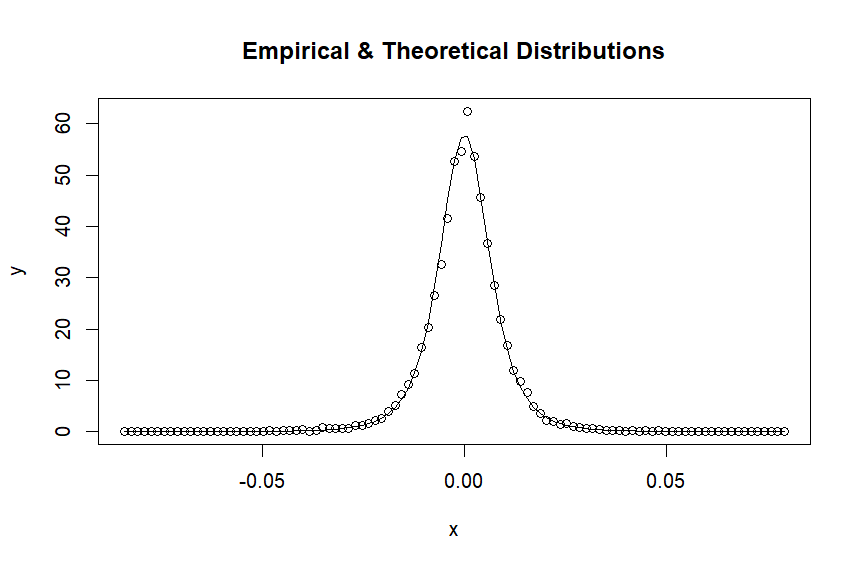}
        \caption{The dots represent the empirical distribution, and the lines the theoretical distribution with $q$=1.43. Out of 22939 truncated data, 313 zeros, about 1.36\%, were discarded. The data was then centered.}
    \label{empirical&theoretical}
\end{figure}

The estimated $q=1.43$. The graph of the empirical distribution of the truncated returns and the $q$-Gaussian density is shown in \cref{empirical&theoretical}.

According to \cite{Hahn2010} with slight changes, $q$-Gaussian density can be expressed by
$$G_q(\beta;z)={\sqrt\beta\over C_q}e_q^{-\beta z^2}=\int_0^\infty {1\over \sqrt{2\pi} v}\exp\left({- z^2\over 2v^2}\right)f_V(v;q)\, dv, \eqno(2)$$ where $$f_V(v;q)\,
=C_{V,q}\beta^{(q-3)/(2(q-1))}\exp\left(-{1\over 2(q-1)\beta v^2}\right) v^{-{2\over q-1}},$$ with \quad $\displaystyle{C_{V,q}^{-1}=\Gamma\left({3-q\over
2(q-1)}\right)\cdot {1\over 2}\cdot [2(q-1)]^{3-q\over 2(q-1)}}$, is the mixing density, and $\Gamma(\cdot)$ is the Gamma function. The independent variable $v$ in $f_V(v;q)$ is the standard deviation of the normal density, possibly an abuse of the notation.

\begin{figure}
\centering
    \includegraphics[scale=.5]{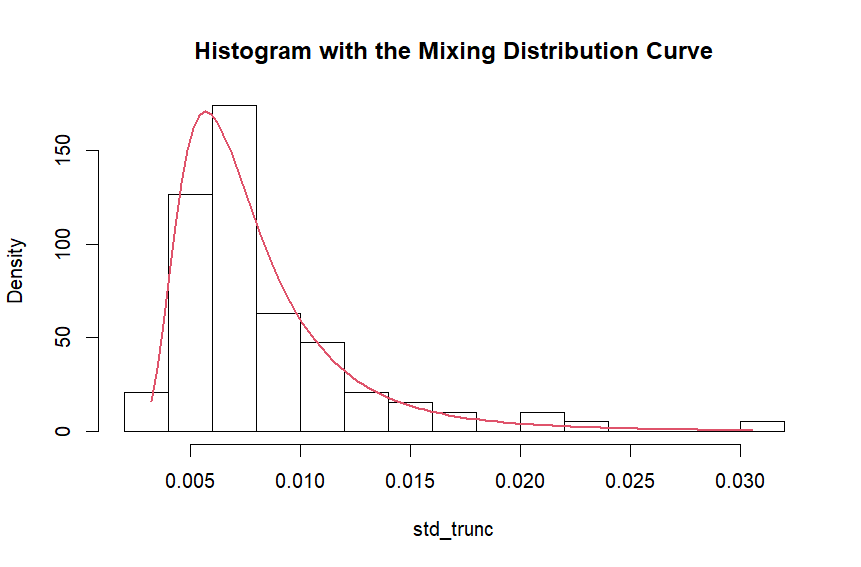}
        \caption{The histogram is the daily standard deviation for each year. It's the same as \cref{hist_sd}. The curve is the theoretical mixing distribution in the $q$-Gaussian. It's the distribution of $1/a$ where $a$ is the square root of a $\chi^2$ random variable.}
    \label{mixing}
\end{figure}

This mixing density is a theoretical distribution to model standard deviations of the truncated returns. Using estimated $q$ and $\beta$, a graph with both the mixing density and sample standard deviation of the truncated returns is shown in \cref{mixing}.

\subsection {Outliers and Negative Binomial Distribution}
\par
As mentioned above, due to overdispersion, Poisson distributions might not be appropriate to model outlier counts. Instead, we use a negative binomial distribution. In general, the distribution is used to model the number of failures in Bernoulli trials, with the probability of success $p$, when the first of $\gamma$ successes occurs. According to \cite{Kagraoka2005}, the first application of the negative binomial model in economics was by \cite{Hausman1984}.
\par
It seems intuitively surprising that a negative binomial distribution is involved here. However, it is known that a negative binomial distribution is also a mixture of Poisson distributions, where the mixing distribution of the Poisson rate $\lambda$ is a gamma distribution. We might imagine that when a Poisson event occurs (or right before it occurs), a $\lambda$ is selected from a gamma distribution. Hence, for a large value of $\lambda$, it is more likely to have a bigger count of outliers.
\par
\begin{figure}
\centering
    \includegraphics[scale=.35]{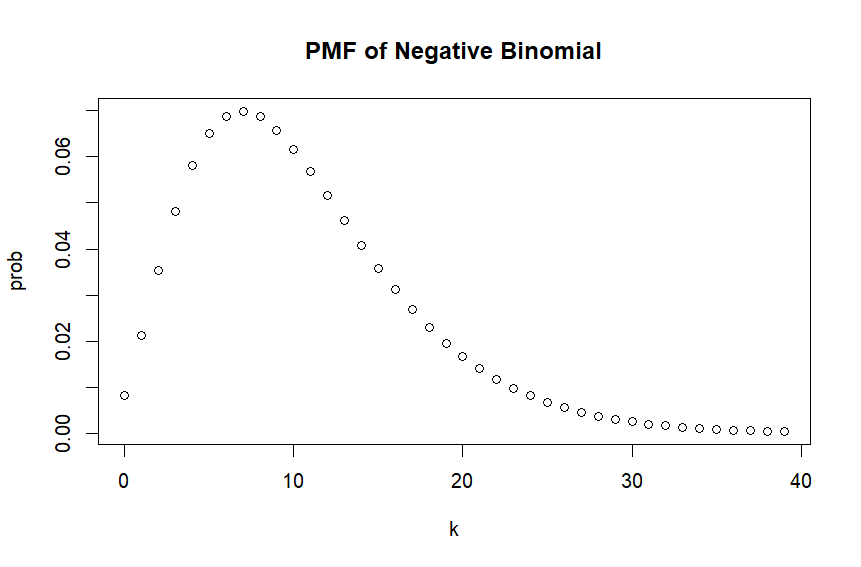}
    \includegraphics[scale=.35]{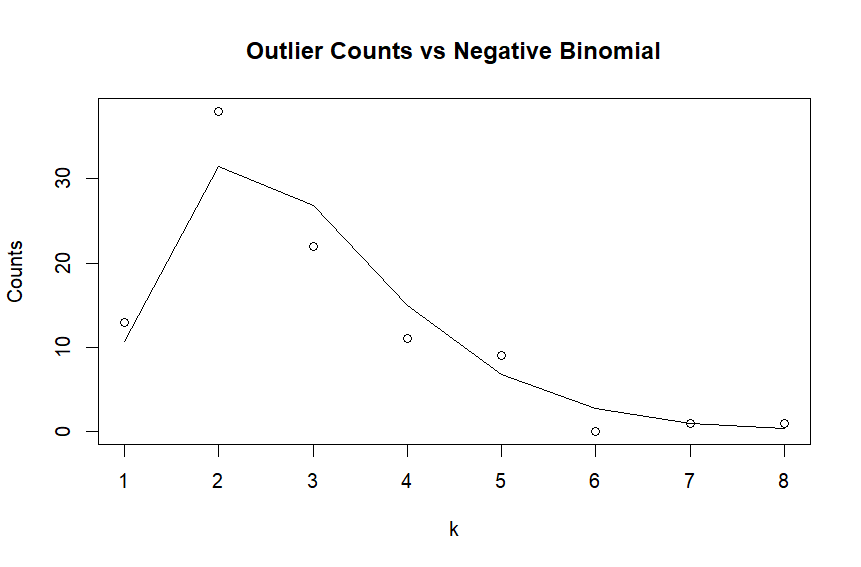}
    \includegraphics[scale=.4]{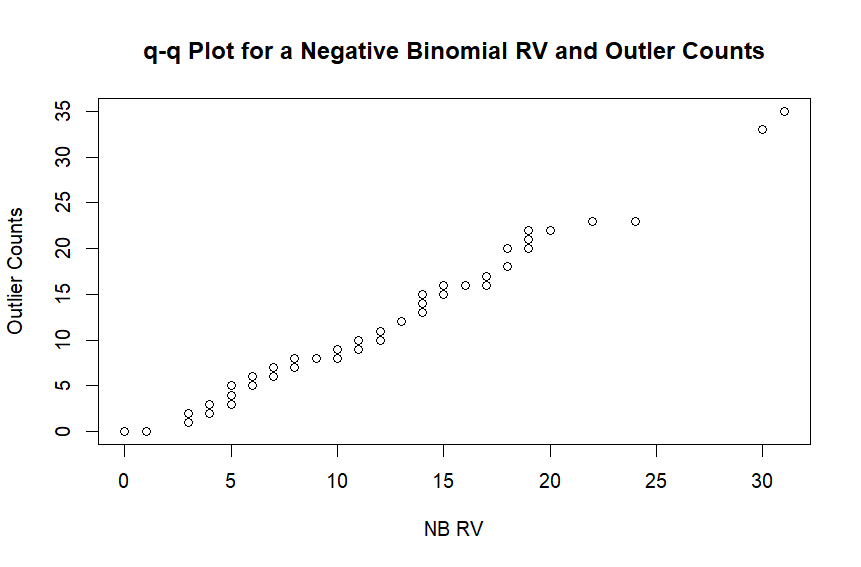}
    \caption{The top left side shows the pmf of negative binomial with $p=0.244$ and $\gamma=3.4$. For the top right, the circles are the counts for the outlier counts grouped in intervals [0,5), [5,10), etc. The line is the same as the left side with counts between [0,5), [5, 10), etc. grouped. The bottom shows the q-q plot for a computer-generated negative binomial rv and the real outlier counts.}
    \label{nb}
\end{figure}

The probability mass function (pmf) of such a negative binomial distribution $NB(\gamma,p)$ is $$ C(k+\gamma-1, k)(1-p)^kp^\gamma.$$ Generalize it to non-integer values, we use the following version: $${\Gamma(k+\gamma)\over \Gamma(k+1)\Gamma(\gamma)}(1-p)^k p^\gamma.\eqno(3)$$ Its expectation is $\gamma(1-p)/p$ and its variance is $\gamma(1-p)/p^2$. Using the method of moments, the estimated $p$ and $\gamma$  are given by $$\hat{p}={\rm{sample\ mean}\over \rm{sample\ variance}},\ \ \ \hat{\gamma}={{\rm{sample\ mean}}\cdot \hat{p}\over 1-\hat{p}}.$$ Our estimates are $\hat{p}=0.244$, and $\hat{\gamma}=3.4.$ The $\gamma$ is annual because the counts of the outliers are annual. The pmf is shown on the top left side of \cref{nb}. Notice the similarity between this and the histogram of the counting numbers of the outliers, the right side of \cref{outlierdist}. The top right of \cref{nb} shows two graphs together.

\par
A q-q plot for counting numbers of outliers and computer-generated $NB(\hat {\gamma},\hat{p})$ random variable is shown on the second row of \cref{nb}.
\par
For the mixing gamma distribution, we adopt the following density function:
$$f_\Lambda(\lambda;s,\tau)={\lambda^{s-1}e^{-\tau\lambda}\tau^s\over \Gamma(s)}\ \ \ \ \lambda>0, s>0, \tau>0.\eqno(4)$$
\par
Here, $s=\gamma$ is the shape parameter and $\tau=p/(1-p)$ is the rate parameter (inverse of the scale parameter). $\lambda$ is the mean for Poisson.
\par
The estimated shape $s=\hat{\gamma}=3.4$, and the rate $\tau={\hat{p}/(1-\hat{p})}=0.3227.$ They are annual data. For daily data, $s=\hat{\gamma}/252=0.0135$ and $\tau$ is again 0.3227.

\section {Stock price dynamics and option pricing formula}
\par

We will assume that the stock price $S_t$ follows the random process
$$ {dS_t\over S_t}=\mu dt+ \sigma dB_t+(J-1)dN(\lambda t), \eqno(5)$$ where $\mu$ is the drift rate, $\sigma$ is the volatility randomly selected from density $f_V(v;q)$ in (2), $B_t$ is the Brownian motion, $J$ is the jump size as a multiple of stock price, and $N(\lambda t)$ is the Poisson process representing the number of jumps that have occurred up to time $t$ with $\lambda$ randomly selected from a Gamma distribution $f_\Lambda(\lambda)$ in (4). All the processes and random variables are considered to be independent of each other. The distribution of $J$ is assumed to be log-normal with density $m\exp(-\nu^2/2+\nu N(0,1))$ where $N(0,1)$ is the standard normal distribution, $m$ is the average jump size, and $\nu$ is the standard deviation of jump size.
\par
When $\sigma$ and $\lambda$ are scalar, this random process becomes Merton's jump-diffusion model (1). When $dN(\lambda t)=0$ and $\sigma$ is a random variable, the process becomes a variance mixture of Brownian motions. In \cite{Hahn2010}, it's called $q-VM$ Brownian motion. We point out that a variance mixture of Brownian motions has a different dynamic structure than that in \cite{Borland2002}, where $dS_t/S_t=\mu dt+\sigma d\Omega_t$ with $\sigma$ scalar and $\Omega_t$ evolving according to a statistical feedback process $d\Omega_t=P(\Omega_t)^{(1-q)/2}dB_t,$ and the evolution of the probability distribution $P$ is nonlinear according to the nonlinear Fokker-Plank equation. They both have continuous paths and $q$-Gaussian distributions as marginals. See \cite{Hahn2010} for more details.
\par
To give the call option formula, let $S$ be the spot price, $K$ the striking price, $T$ the time to maturity, and $r$ the interest rate (assumed to be constant for simplicity). Also, we assume no dividend. Then the European call option price is given by

$$P_{GJD}(S,K,r,T)=\int_0^\infty \sum_{k=0}^\infty {e^{-m\lambda T}(m\lambda T)^k\over k!}P_{VMBM}(S,K,r_k, T)f_\Lambda(\lambda;s,\tau)\, d\lambda,\eqno(6)$$
where $$P_{VMBM}(S,K,r_k,T)=\int_0^\infty P_{BS}(S,K, v_k, r_k, T)f_V(v;q)dv,$$
with $v_k=\sqrt{v^2+k\nu^2/T}$ and $r_k=r-\lambda(m-1)+k\log(m)/T$\cite{Merton1976}\cite{Peter2011}. Here $GJD$ means generalized jump-diffusion process, $VMBM$ means variance mixture of Brownian motions, $P_{BS}$ is the Black-Scholes formula, and $\log$ is the natural logarithm.

\section{Empirical results using SPY}
\par
SPY closing data was taken from Google.com using R package quantmod. The starting date was 2007/1/4. As in the previous discussion, the first $17\times 252$ closing points were selected to estimate some statistics. The results (daily) are in the following table:

\begin{table}[h!]
\centering
\begin{center}
\begin{tabular}{|c|c|c|c|c|c|c|c|}
  \hline
  $q$ & $\beta$ & $\hat{\gamma}$ & $\hat{p}$ & $s$ & $\tau$ & $m$ & $\nu$ \\
  \hline
  1.4 & 14582.54 & 0.029 & 0.35 & 0.029 & 0.54 &0.9923& 0.03777  \\
  \hline
\end{tabular}
\end{center}
\caption{Parameters of the model.}
\label{table:1}
\end{table}

 \par
In Table \ref{table:1}, $q$ and $\beta$ are the parameters in the $q$-Gaussian (2), $s$ and $\tau$ in the Gamma distribution (4), $m$ and $\nu$ in the lognormal distribution for jump sizes. They are used in the pricing formula (6). $\hat{\gamma}$ and $\hat{p}$ are the parameters in the negative binomial distribution (3), which are not used directly in (6).
All the parameters are daily. \footnote{The average jump size $m$ for 95 year S\& P500 records is 0.9962895, and the standard deviation $\nu$=0.03687745. Since $m$ is close to 1, the generalized jump-diffusion model (5) seems to be close to the variance mixture of Brownian motions.}
\par
The risk-free interest rate is 4\% annually, about 0.011\% daily.
\par
We chose the starting date to be 2/5/2024 since it was the day this paper was edited. It was a Monday. Data in Table \ref{table:2} below were downloaded from TD Ameritrade at 4:19 PM EST. 

\begin{table}[h!]
\centering
\begin{center}
\begin{tabular}{|c|c|c|c|c|}
  \hline
  Date & Bid & Ask & Average of Bid-Ask & Close \\
  \hline
  2/5/2024 & 492.42 & 492.45 & 492.435 & 492.55  \\
  \hline
\end{tabular}
\end{center}
\caption{Closing data for S\&P500.}
\label{table:2}
\end{table}
\par
It's not clear whether the closing price should be used for computing option prices. For example, the 1-day call option with a strike price of 475 (ITM) was closed at 16.50, which was lower than the SPY closing price of 492.55 minus the strike price of 475, which is 17.55. Similarly, the 1-day call option with a strike price of 477 was closed at 15.39, again, smaller than the close minus the strike. However, if we look at the averages of bid and ask for option prices, the values are more in line. So, for option prices at different maturities, the bid-ask averages will be used here. Similarly, $S$ is chosen to be the average of bid and ask for SPY, which is 492.435, or 492.44, rather than the closing price of 492.55.
\par
Another issue is how to count weekend days. Looking at theoretical option prices and real market option prices, ignoring weekend days and counting only work days would give better fits. Hence, we will discard weekend days.
\par
\begin{figure}
\centering
    \includegraphics[scale=.35]{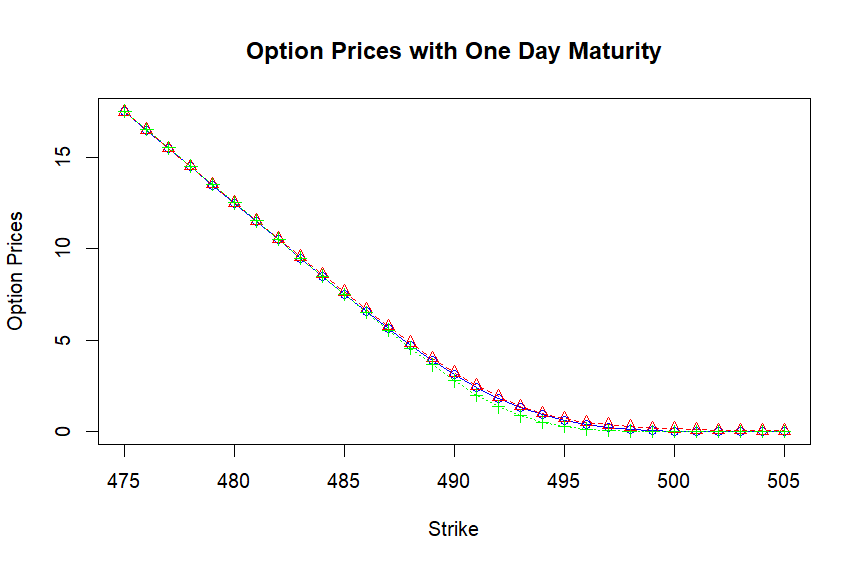}
    \includegraphics[scale=.35]{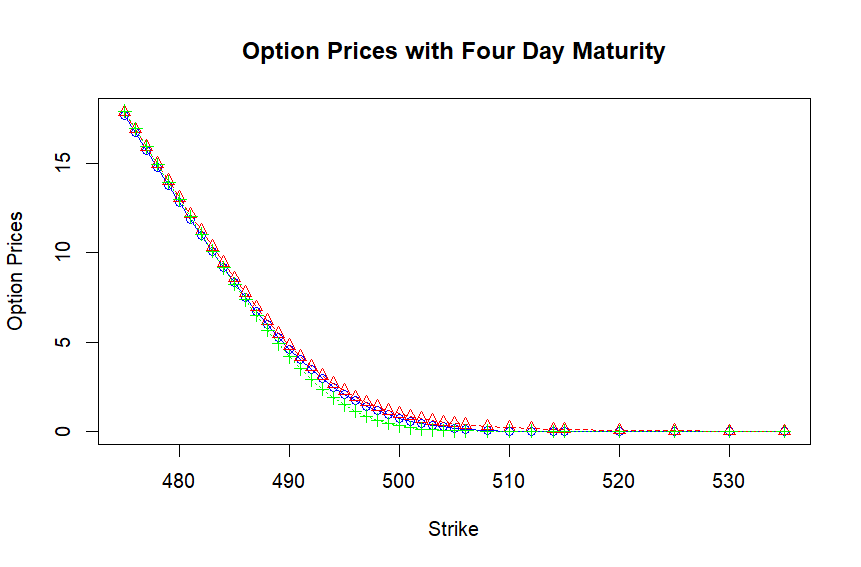}
    \includegraphics[scale=.35]{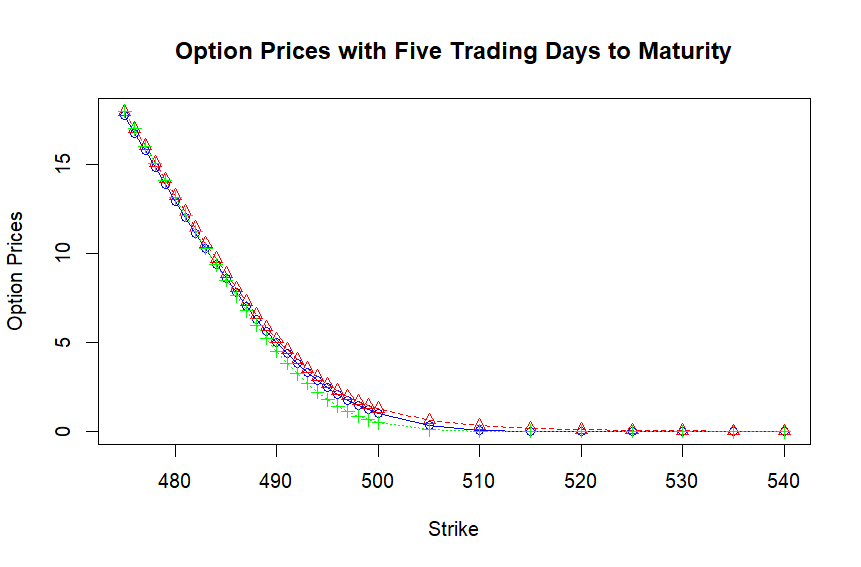}
    \includegraphics[scale=.35]{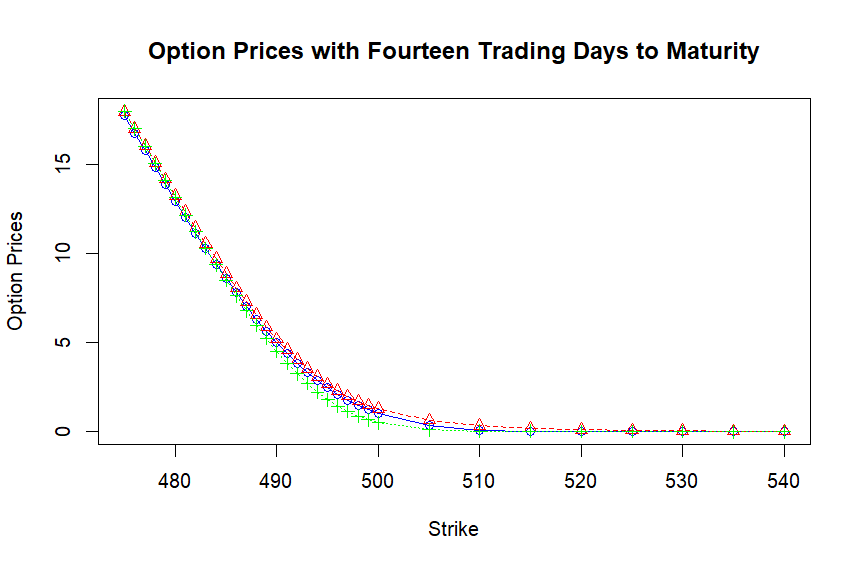}
    \includegraphics[scale=.35]{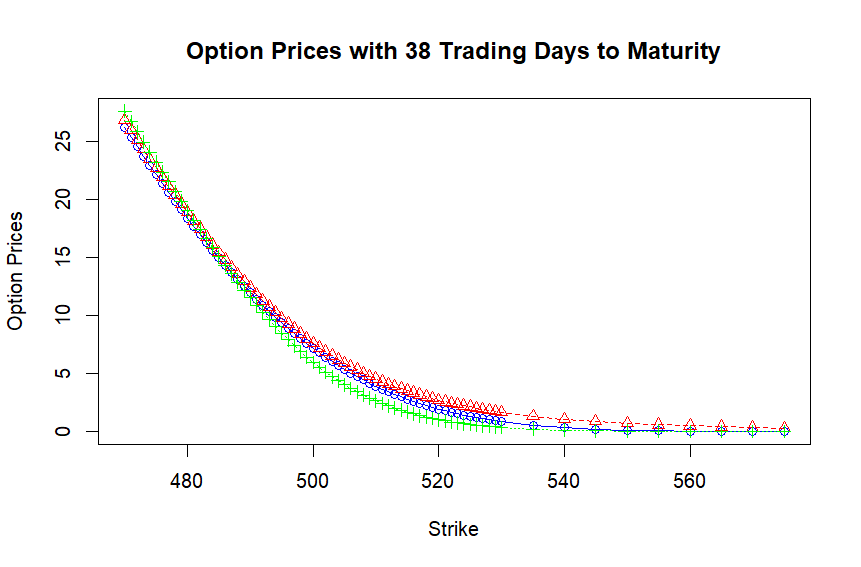}
    \includegraphics[scale=.35]{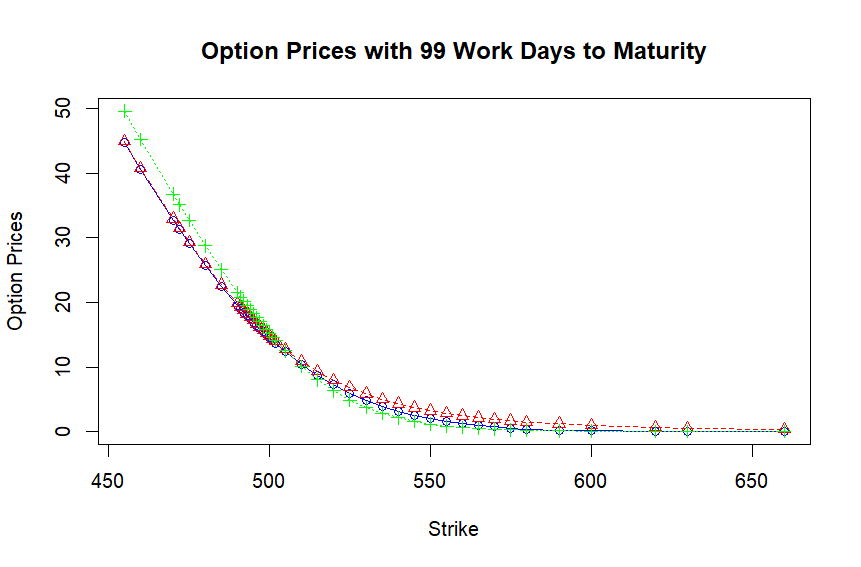}
    \caption{The green lines with crosses are market prices. The blue lines with circles are prices obtained by the Black-Scholes model. The red lines with triangles are prices obtained by the generalized jump-diffusion model.}
    \label{optionprices}
\end{figure}

\begin{figure}
\centering
    \includegraphics[scale=.35]{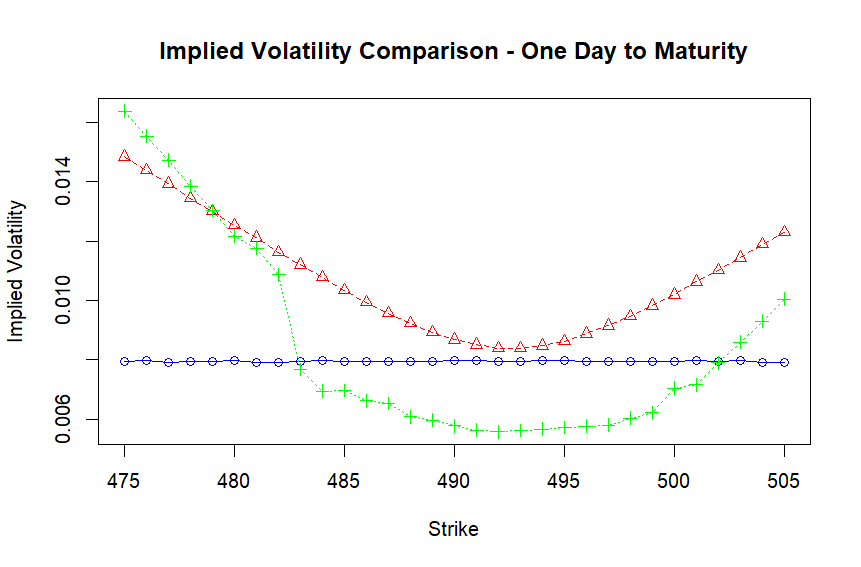}
    \includegraphics[scale=.35]{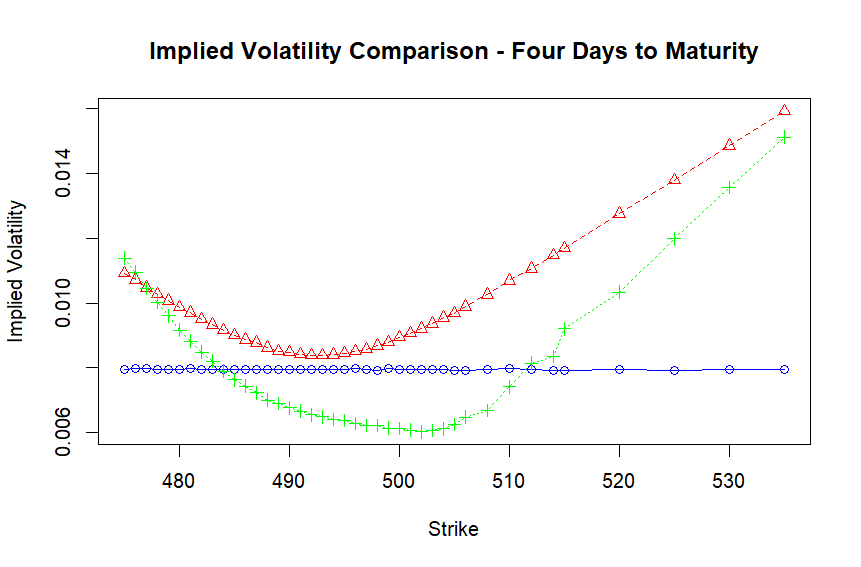}
    \includegraphics[scale=.35]{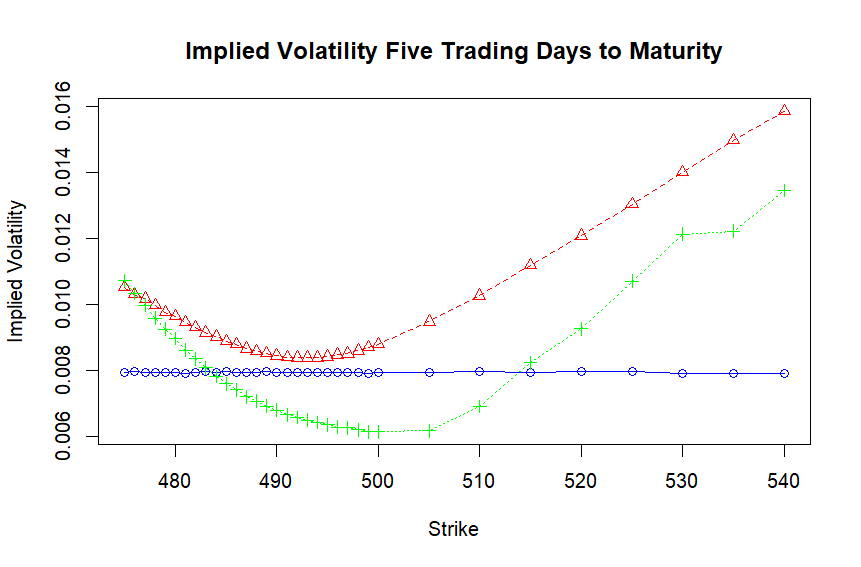}
    \includegraphics[scale=.35]{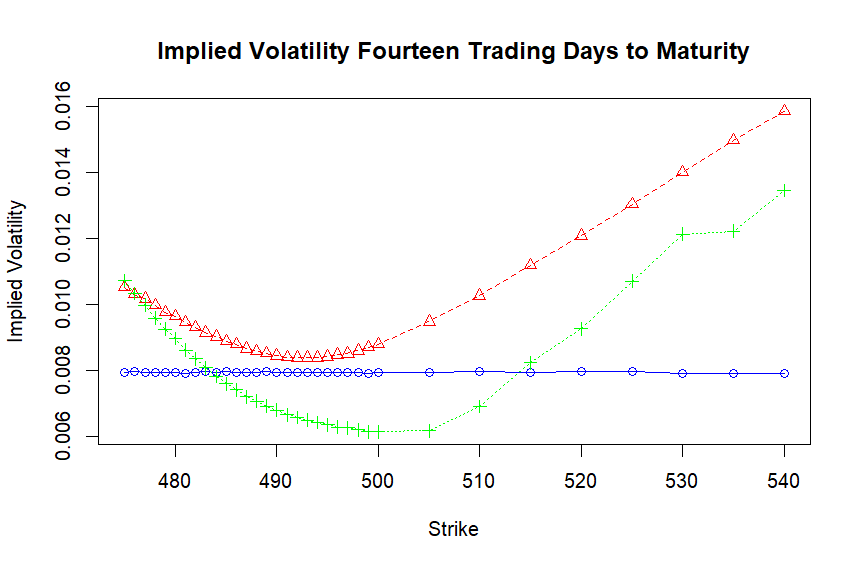}
    \includegraphics[scale=.35]{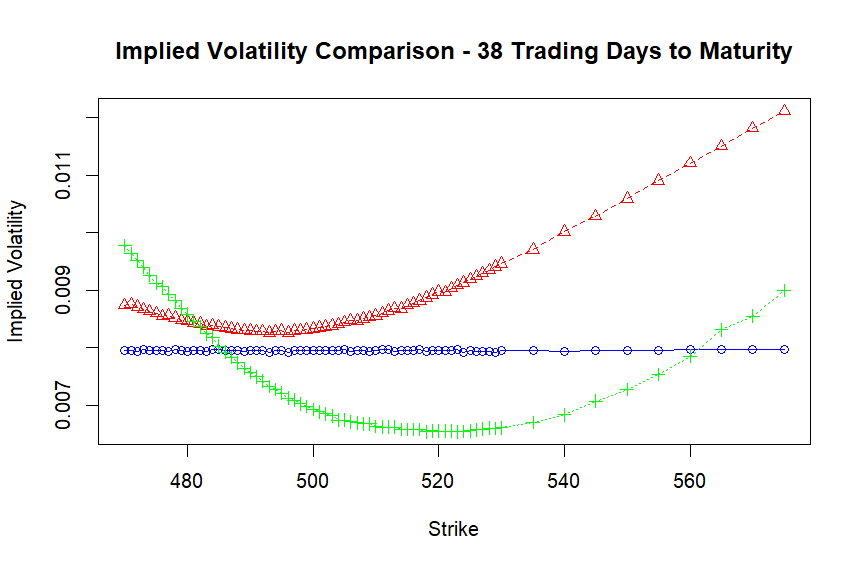}
    \includegraphics[scale=.35]{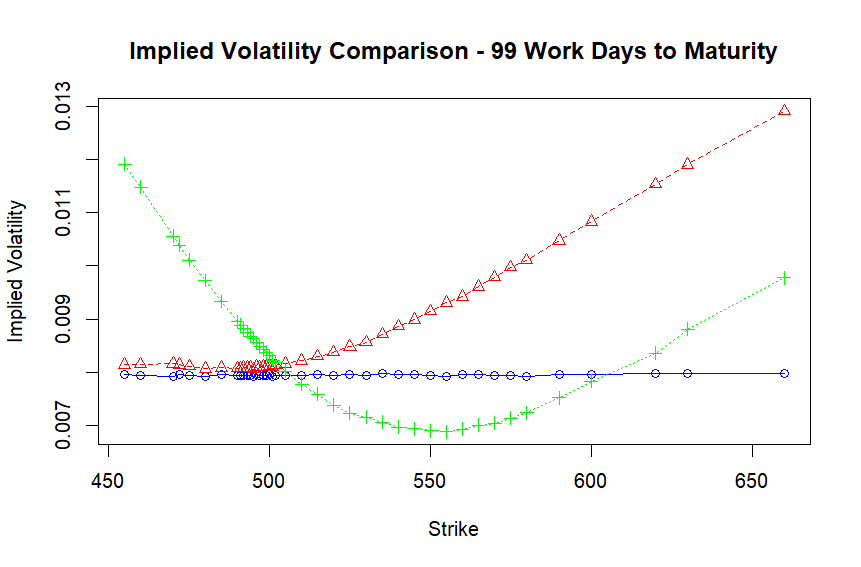}
    \caption{The green lines with crosses are implied volatilities for market prices. The blue lines with circles are for prices obtained by the Black-Scholes model. The red lines with triangles are for prices obtained by the generalized jump-diffusion model.}
    \label{ivs}
\end{figure}

 
Call option prices for 1/1 day (the first number is for the work day and the second for calendar day), 4/4 days, 5/7 days, 14/18 days, 38/52 days and 99/137 days to maturity are recorded in \cref{optionprices}. Their implied volatilities are shown in \cref{ivs}.

\par 
From these results, we may claim the following:

(1) The prices obtained from the generalized jump-diffusion(GJD) model are only slightly higher than those obtained from the Black-Scholes model. This is reasonable since the jump size $m$ is close to 1. However, the implied volatility smiles are clearly shown for the prices obtained from the GJD model. So are the smiles shown for the market option prices.

(2) Compared with the market prices, the prices obtained from the two models overestimated option prices around the spot price (``at the money") when the time to maturity is relatively short, but underestimated the prices when the time to maturity is relatively longer (100 trading days or longer). They always overestimate the prices for options deep out of the money. When options are deep in the money, they underestimate prices, with differences higher when the time to maturity is longer. 

(3) The implied volatilities for option prices obtained from the GJD model reach the smallest when they are at the money and symmetric to the smallest volatilities. However, the smiles obtained by market option prices shifted to the right more when the time horizon was longer.

\section{Conclusions and discussions}
\par 
We presented somewhat strong evidence that the distribution of the US stock market returns (indices) might be modeled by variance mixtures of normals ($q$-Gaussians) after the returns are truncated. For extreme returns, negative binomial distributions might be good candidates. Based on this evidence, a generalized jump-diffusion model was proposed and an explicit formula for call options was obtained. The option prices obtained from the model, together with the prices obtained from the Black-Scholes model, were compared with market data. The option prices obtained from the GJD model do give implied volatility smiles. However, they do not match the smiles for real-market option prices very well. 

From the implied volatility smiles resulting from real option prices, we might conclude that the market participants on 2/5/2024 expected the future volatilities to be less than the historical volatility. This expectation was generally correct since the following 5-10 days, the volatilities were indeed smaller than the historical volatility we used in the BS model and GJD model. However, we might not be able to conclude that the implied volatilities obtained from the real market option price data can always predict future volatilities. Maybe partially, because when there is no foreseeable scheduled news causing large market volatility, the market is more likely to have smaller volatility. But nobody would be able to predict unforeseeable events that could cause the market to have large volatility. The GJD model always considers the possibility of extreme events. Hence, when the market has been stable for a while,  the implied volatility from market option price data tends to be smaller than what the BS model and GJD model predict. However, right after a market crash, the opposite might happen. This is consistent with the comparison between the pre-1987 market crash and post-crash \cite{CHRISTENSEN1998}.

\bibliographystyle{plain} 
\bibliography{references}

\end{document}